# Multi-Scale Hourglass Hierarchical Fusion Network for Single Image Deraining

Xiang Chen[1]  Yufeng Huang[1*]  Lei Xu[2]

1 College of Electronic and Information Engineering, Shenyang Aerospace University, Liaoning, China
2 Shenyang Fire Science and Technology Research Institute of MEM, Liaoning, China

[1] cv.xchen@gmail.com, yufengh_sau@126.com, [2] syfri_xulei@126.com

## Abstract

*Rain streaks bring serious blurring and visual quality degradation, which often vary in size, direction and density. Current CNN-based methods achieve encouraging performance, while are limited to depict rain characteristics and recover image details in the poor visibility environment. To address these issues, we present a Multi-scale Hourglass Hierarchical Fusion Network (MH²F-Net) in end-to-end manner, to exactly captures rain streak features with multi-scale extraction, hierarchical distillation and information aggregation. For better extracting the features, a novel Multi-scale Hourglass Extraction Block (MHEB) is proposed to get local and global features across different scales through down- and up-sample process. Besides, a Hierarchical Attentive Distillation Block (HADB) then employs the dual attention feature responses to adaptively recalibrate the hierarchical features and eliminate the redundant ones. Further, we introduce a Residual Projected Feature Fusion (RPFF) strategy to progressively discriminate feature learning and aggregate different features instead of directly concatenating or adding. Extensive experiments on both synthetic and real rainy datasets demonstrate the effectiveness of the designed MH²F-Net by comparing with recent state-of-the-art deraining algorithms. Our source code will be available on the GitHub: https://github.com/cxtalk/MH2F-Net.*

## 1. Introduction

Rain streaks undesirably degrade the scene visibility and severely impair the performance of subsequent vison tasks, as object detection [1], recognition [2] and tracking [1]. Images captured under rainy situations significantly undergo degradation, which certainly subject to detail blurring, color distortion, and content obstruction. Single image deraining has thus attracted increasing attention and become a vital preprocessing step in the practical applications.

[*] Corresponding author

Generally, the ultimate goal of deraining is restore the clean background $B$ from its observed rainy image $O=B+R$ with a rain component $R$. Due to the lacking information of background and rain layers, single rain removal can be served as a highly ill-posed problem. How to solve the ill-posed solution and stably obtain the unique rain streaks layer become a core topic for deraining. Existing deraining researches are mainly classified into the model-based and data-driven methods. Model-based approaches generally treat the deraining as an optimization problem and rely on the prior knowledges to analyze the image layers. The typical approaches are driven by the following theories: image decomposition [3], Gaussian Mixture Model (GMM) [4], Discriminative Sparse Coding (DSC) [5] and filter-based model [6]. The model-based methods apply handcrafted prior assumptions on the rain streaks and background, and only perform well to certain conditions.

The other popular solution is data-driven methods that serve the rain removal as learning procedure of non-linear function and explore the appropriate parameters to extract the rainy layer from background scene [7]. Motivated by the learning-based breakthrough, many data-driven learning methods emerge and gain outstanding deraining performance. Deep CNN based researches exploit the automatically way to extract effective features, like rain masks [8] and background scenes [9]. Through the architecture design, a series of work employs the recursive structure [10], or recurrent network [8] to eliminate the rain streaks progressively. Generative Adversarial Networks (GAN) based methods introduce the generator and discriminator to restore the rainy image to a clean one. The CGAN [11] is used to get better feature distribution, in order to suppress the domain differences between the generated result and real ground truth. In [12], depth-guided GAN is applied to compensate the details lose and reduce the incoming artifacts. The semi/unsupervised deraining methods [12, 13] try to improve the network capability and generality by learning from real rainy images. Although achieve great progress, the above deraining models have deficiencies in fully using of multi-scale rain features and restoring structure details. Few attempts have been exploited to preserve the spatial features and eliminate the redundant information.

In this paper, we develop a novel Multi-scale Hourglass

Hierarchical Fusion Network (MH²F-Net) for single image deraining, which consists of multi-scale feature extraction, hierarchical feature distillation and feature aggregation. To be specific, a multi-scale hourglass is involved in the feature extraction stage, that can extract rich detail features through the integrated dense connected residual blocks and multi-scale parallel structure. Then, hierarchical feature distillation employs the dual attention mechanism to adaptively recalibrate the feature responses and remove unnecessary features. In addition, residual projected feature fusion aims to maximize the use of features from different sources and fuse information recurrently.

In summary, our main contributions are summarized as:

1 We propose a novel Multi-scale Hourglass Extraction Block (MHEB) to extract features from different scales and take full advantages of these local and global features to improve the feature extraction capability.

2 To our knowledge, Hierarchical Attentive Distillation Block (HADB) is first constructed to recalculate the feature map of rain streaks. A hierarchical distillation structure is used to avoid feature disappear and reduce useless features. Adding the dual attentive distillation mechanism, the distillation structure further helps to preserve the spatial and contextual information for better feature recalculation.

3 To better characterize rain features from different sources, Residual Projected Feature Fusion (RPFF) is designed to improve the network discriminative ability and gradually aggregate different feature maps to produce realistic derained results.

## 2. Related Work

As a highly ill-posed problem, single image deraining has drawn increasing attention in the past few years. We briefly review the most related deraining technologies and multi-scale learning strategies as follows.

**Model-based methods** are designed by describing the rain characteristics with the handcrafted image features, or constraining the ill-posed solution using prior knowledge. [3] first proposes the single image decomposed problem and eliminate the rain streaks in the high frequency part via the morphological component analysis-based dictionary learning way. [5] designs a discriminative sparse coding to approximate the background and rain layers. In [4], GMM based patch prior is employed to model the rain streaks and background separately by accommodating multiple scales and orientations.

**Data-driven methods** consider the deraining process as a non-linear mapping function and apply various ways to automatically figure out the rain streaks and background. Many data-driven learning approaches realize a significant performance boost for single image deraining and the most related researches are summarized in the following part. Fu et al [14] first tries to take away the rain streaks by using multi-layer CNNs and then advance the process via deep detail Network [15]. In [16], RESCAN introduces the dilated convolution to get contextual information, and applies recurrent neural network-based way to remodel the rain feature. SPANet [17] designs the recurrent network to capture the spatial contextual information in a local-to-global manner. [13] utilizes a semi-supervised learning method to calculate the residual difference between the input image and the derained image. RCDNet [18] uses a convolution dictionary to represent the rain feature and simplify the network with proximal gradient descent technology. [19] offers a DRD-Net, which has two-branch parallel structures to remove the rain streaks and repair the detail.

**Multi-scale learning strategy** can bring an in-depth view of image layout and strengthen the feature extraction capability to certain extent [7]. The rain pattern displays the obvious self-similarity, so that the correlated messages across different scales may help to increase the feature representation ability. Few attempts have been exploited to study the multi-scale learning for single image deraining. [9] employs a lightweight pyramid network to construct a series of parallel subnetworks to estimate the rain features separately. A residual cascaded pyramid network [20] is designed to alleviate the difficulty of rain image decomposition in a coarse-to-fine process. In [21], a recurrent hierarchy network is introduced to explore the correlation relationship between neighboring stage progressively. The work of DCSFN [22] explores the cross-scale manner and inner-scale fusion to deal with the rain removal. [23] applies the multi-scale progressive fusion structure to fuse the feature and boosts the end-to-end deraining training.

## 3. Proposed Method

In this section, we briefly introduce the overall framework of designed MH²F-Net for single image deraining. The key modules of designed network are illustrated in the following subsections, as well as a description of loss function.

### 3.1. The Overall Structure of MH²F-Net

We propose an end-to-end framework to restore derained images using Multi-scale Hourglass Hierarchical Fusion Network (MH²F-Net), which is comprised of MHEBs and a HADB for feature extraction and distillation respectively, as well as the RPFF to progressively aggregate information. As shown in Figure 1, MH²F-Net can be divided into three stages: feature extraction, distillation, and aggregation.

In the feature extraction stage, we apply a 3 × 3 convolutional layer to extract low-level original features $L_o = F_{3 \times 3}(I_{rain})$, and is also used as the first MHEB input. Then, the original feature maps pass through the Stacked Hourglass Group (SHG), which consists of $N$ MHEBs, yielding deep features as $L_e = F_{SHG}(L_o)$.

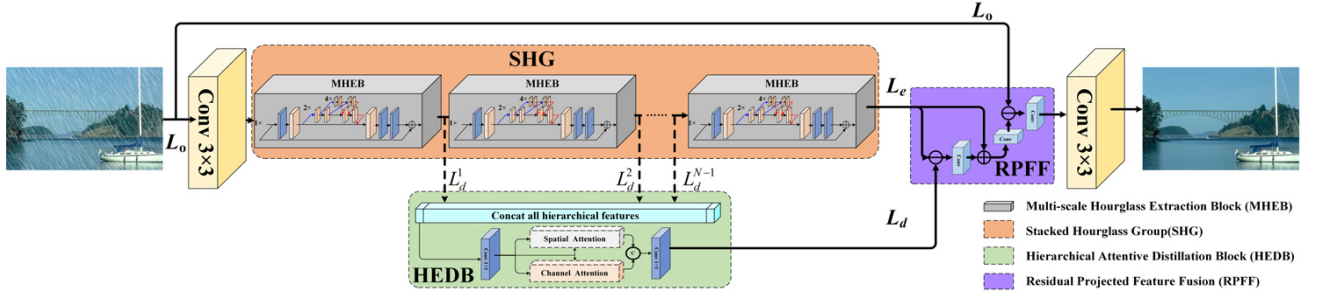

Figure 1. The overall framework of our designed Multi-scale Hourglass Hierarchical Fusion Network (MH$^2$F-Net). The orange, green and purple blocks represent the feature extraction, distillation and fusion module, respectively.

The feature distillation stage employs a HADB to maximally use of hierarchical features and remove dispensable features. The feature distillation process can be defined as:

$$L_d = F_{HADB}([L_h^1, L_h^2, \cdots, L_h^{N-1}]) \quad (1)$$

where $F_{HADB}$ represents the HADB process and $L_d^n$ (n=1, 2, …, N-1) denotes the output of the n-th MHEB, and $L_d$ is expressed as the distilled hierarchical feature.

After extracting and distilling stages, we can generate the original, extracted and distilled feature maps independently. Therefore, RPFF is then utilized to aggregate the features from three branches for aggregating exactly feature. The final output of derained can be formulated as:

$$I_{derain} = F_{3\times3}(F_{RPFF}(L_o, L_e, L_d)) \quad (2)$$

where $F_{RPFF}$ denotes the RPFF mechanism. $L_o$ is the original feature, $L_e$ represents the extracted deep feature, $L_d$ expresses as the distilled hierarchical features.

### 3.2. Multi-scale Hourglass Extraction Block

Feature extraction is a critical step in the deraining research, as the rain streaks may vary in size, direction with complex spatial and context background. We design a novel MHEB module, in order to fully exploit the multi-scale rain feature extraction.

**The Hourglass Network** [24] was first introduced in the human pose estimation by the demand to capture features at different scales. As shown in Figure 2(a). a multi-scale hourglass network has fully stacked parallel convolution streams that down-sample features into certain scales and then up-sample them back to the original one. So multi-scale hourglass network can effectively capture both local and global features in the multi-scale levels. To be specific, our MHEB contains three parallel streams that the first one deals with the original scale and the other two down-sample the scales as 1/2 and 1/4, respectively. The simple nearest neighbor up-sample with skip connections can help to preserve spatial information at each scale. The multi-scale hourglass network has a symmetric topology structure, so that every scale presents the way of top-down and also has a corresponding layer going bottom-up. In each stream, the **Densely Connected Residual (DCR) block** [25] is employed as the basic unit to deeply guide the rain streaks extraction. DCR applies the Residual Net to transport the feature deeper, and also utilities the DenseNet to share the feature information to all the following layers.

Our feature extraction stage stacks multi-scale hourglass blocks, allowing the feature reevaluation across the scales and blocks. The extraction output is an interaction of many multi-scale features that should be used to form a coherent understanding of the rain streaks pattern.

### 3.3. Hierarchical Attentive Distillation Block

For the deraining issue, it is critical to fully exploit the rain features and transfer them to the end for rain removal process. As the depth of network increases, the spatial expression capability gradually decreases during transmission, and massive redundant features aimlessly generate. How to solve these problems will directly affect the deraining quality. In this work, we employ a simple hierarchical feature distillation structure with attention mechanism termed Hierarchical Attentive Distillation Block (HADB), as seen in Figure 3. The outputs of each MHEB have been served as the inputs of HADB gradually, in order to avoid the spatial information loss during the feature transformation pooling and transmission. The core of HADB emphasis on recalibrate dual attention feature responses to realize the feature distillation.

**Dual attentive distillation mechanism** is used to eliminate massive aimless features and extract more useful hierarchical features. Inspired by the success of attentive mechanism in vision fields, we utilize the dual attention unit to extract features in the hierarchical distillation block. The dual attention mechanism explores the useful spatial and channel components and only allows informative features to pass further. So the feature recalibration is realized by using spatial attention [26] and channel attention [27] mechanisms. In the spatial attention step, a global average and max pooling process carries out to capture the inter-spatial dependencies of different features.

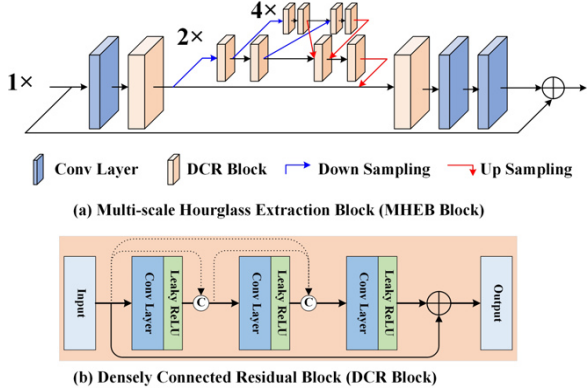

Figure 2. The structure of feature extraction. (a) Multi-scale Hourglass Extraction Block and (b) Densely Connected Residual Block.

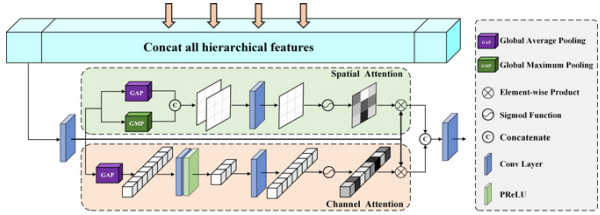

Figure 3. The architecture of Hierarchical Attentive Distillation Block (HADB) for feature distillation stage.

The channel attention is involved to exploit the feature contextual information and mine the relationship between hierarchical features. Motivated by autoencoder, the bottleneck layers (1×1) are applied at the head and end of the attention module, that can help to achieve the feature distillation.

### 3.4. Residual Projected Feature Fusion

As observed in the deraining stage, it is an effective way to get rain streaks by using multi-scale features. It is still unclear about how to effectively fuse the features from different sources into a comprehensive feature map. Existing methods [9, 21] usually deal with each scale separately or simply add them together. Different from the mentioned methods, our proposed network explores a novel RPFF to gradually combine features together, as shown in Figure 4. The feature extraction and distillation structures of $MH^2F$-Net produce three feature maps independently as the original ($L_o$), extracted ($L_e$) and distilled ($L_d$) feature map. In order to concentrate on more informative feature maps, we get the residual $R_{ed}$ by calculating the difference between $L_e$ and $L_d$. Further the residual $R_{ed}=L_e-L_d$ is processed by the convolution process and then is added back to obtain $F_{ed}$.

$$F_{ed} = conv(R_{ed}) + L_e \quad (3)$$

where $R_{ed}$ denotes the features existing in one source but lacking in the others. Such residual projected procedure can greatly improve the discriminative ability, that assists the network to concentrate on the distinct features while bypassing the common parts. We adopt the back-projection strategy to incorporate distillation information to regularize the rain streaks feature extraction and obtain the derained image. The final fused feature $L^*$ is computed by,

$$L^* = conv(L_o - conv(F_{ed})) \quad (4)$$

where $L_o$ is the original feature map, *conv* is a convolution.

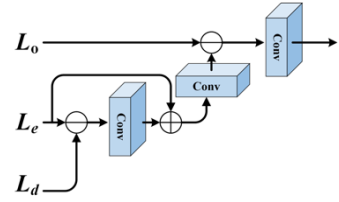

Figure 4. The flow of Residual Projected Feature Fusion and its inputs are the original (Lo), extracted (Le) and distilled (Ld) feature maps.

The RPFF module guarantees an effective rain feature learning through dynamically combining different feature extraction branches, improving the discriminative feature capacity compared with directly adding or concatenating. Meanwhile, it can also avoid the over-deraining or under-deraining caused by the global residual learning.

### 3.5. Loss Function

Generally, the derained output should be close to the ground truth in certain level. To train our designed network, a hybrid loss function is employed, consisting of structural similarity index (SSIM) [28] loss and $L_1$-norm loss [29]. Specifically, the SSIM loss is applied to evaluate the structural similarity, which can better preserve high-frequency structure information. $L_1$-norm loss offers an effective way to constrain the differences between the color and luminance characteristics.

These two loss functions can be formulated as:

$$L_1 = \frac{1}{N}\sum_{i=1}^{N} \|R - GT\| \quad (5)$$

$$L_s = 1 - SSIM(R, GT) \quad (6)$$

where $L_1$ and $L_s$ are defined as the $L_1$-norm and SSIM loss functions, respectively. $R$ is the derained image and GT is the ground truth image.

By combining the weighted sum of SSIM and $L_1$-norm loss, our final hybrid loss function can be expressed as:

$$L_{total} = L_1 + \lambda L_s \quad (7)$$

where λ is a weight parameter, empirically set equal to 0.2.

### 4. Experimental Results

In this section, we briefly introduce the rainy datasets and implementation information. Then, extensive deraining experiments are involved to evaluate the designed MH$^2$F-Net against the recent comparing algorithms on both synthetic and real-world rainy datasets. Finally, ablation studies related to the essential modules are conducted to analyze the effectiveness of our proposed framework.

### 4.1. Experiment Settings

**Datasets setup.** We conduct extensive derained experiments on three updated synthetic datasets: Rain200L/H [8] and Rain1400 [14] including various rain streaks with different patterns. In addition, real-world datasets are selected to estimate the performance of deraining methods. In detail, one is SPA-Data [17] that provides real rainy images and its ground truth is obtained by using multi-frame fusion and human labeling [17], and the other one is from Internet. The comprehensive descriptions of training and testing can be seen in Table 1, including the synthetic and real-world datasets.

**Implementation details.** The overall structure and parameter settings of designed MH$^2$F-Net are illustrated in Figure 1. In MH$^2$F -Net, the number of MHEBs is set to 8 for a better feature extraction. During the training process, the loss weight λ is 0.2 and augment data by randomly cropping 64×64 patch pairs with horizontal flipping. Using the Adam optimization, the parameters are defined as: initial learning rate is $1\times10^{-3}$ and batch size is 16, where $\beta_1$ and $\beta_2$ take the default values of 0.9 and 0.999 respectively. To obtain better performance, we train our model with 200 epochs for the Rain200L/H datasets, 100 epochs for Rain1400 dataset and 25 epochs for SPA-Data dataset. All the training and testing process are performed using PyTorch on a workstation with the NVIDIA Tesla V100 GPU (16G).

### 4.2. Comparison with the State-of-the-Art Methods

The proposed MH$^2$F-Net is compared to other five recent state deraining approaches, including RESCAN [16], SPANet [17], SIRR [13] , RCD-Net [18] and DRD-Net [19]. As the availability of ground truth in certain datasets, the quantitative analysis carries out by using PSNR and SSIM. The qualitative comparisons present several challenging samples both in synthetic and real-world datasets.

**Quantitative Evaluations:** To objectively evaluate the deraining performance, we adopt two classical quality metrics: PSNR and SSIM. Table 2 shows the average evaluation criteria of classical updated synthetic and real-world datasets with the recent comparing deraining methods. Through the Table 2, the proposed model achieved the highest values both in PSNR and SSIM, reflecting the superior generality and robustness of our MH$^2$F-Net in both synthetic and real-world datasets.

Table 1. Descriptions of synthetic and real-world datasets. Values in each column indicate the number of rain-free/rainy image pairs.

| Datasets | Rain200L | Rain200H | Rain1400 | SPA-Data | Internet-Data |
|---|---|---|---|---|---|
| Train-Set | 1,800 | 1,800 | 12,600 | 638,492 | 0 |
| Test-Set | 200 | 200 | 1400 | 1000 | 167 |
| Type | | Synthetic | | Real-World | |

Specially, the greater PSNR scores indicate better capability to remove rain streaks from the rainy images. The higher values of SSIM denote that the derained images are uniformly closer to the ground truths by better restoring the image details. The significantly increased values in Rain200H and Rain1400 reveal that our approach could properly recover derained images, especially in the heavy rain as well as the diverse rainy condition. Though more complex in nature rainy scene, the excellent marks prove our network deraining ability in the real-world conditions.

**Qualitative Comparisons:** For the visual comparison, some challenging examples of synthetic datasets directly show derained differences, as particularly seen in Figure 5. As displayed, the examples contain various rain density with complex background scenarios. Through the qualitative presentation, SIRR, RESCAN and SPANet leave some rain streaks after the deraining process, especially serious in heavy rain cases. Obviously, DRD-Net and RCDNet do a better deraining performance in most rain conditions, while these methods appear to over-smooth the pictures. By checking the zoomed parts, the deficient exposures can be found that they blur the contents and fail to restore the details, and similar defects can also be seen in the above other approaches. In contrast, our developed model can effectively remove majority of rain streaks even in the heavy rain conditions, as well as guarantee the details structure closest to the ground-truth.

To further verify the practical use, we compare the developed model against other competing methods on two classical real-world rainy datasets. Figure 6 presents four real-world examples, as the above two of SPA-Data and the following ones from Internet-Data. Due to the complex distribution of rain streaks in real world, all the competing deraining methods have the rain streaks remain even under fewer and smaller raindrops conditions, as displayed in Figure 6(a). As light changes in Figure 6(b) (first line), all the competing approaches fail to eliminate the rain streaks from complicated brightness surrounding. Zooming the color boxes, our proposed method outperforms other competitors in restoring details and removing rain streaks with light effect in a dark environment. The second line of Figure 6(b) contains various spatial and content scenarios. Owing to some haze existing, the rain streaks far from camera are difficult to figure out. Only RCDNet and ours remove majority of rain streaks. Even in the complex surrounding, it can be observed that our model achieves better derained results against other competing methods in preserving image details and texture information.

Table 2. PSNR and SSIM comparisons on four benchmark datasets. Bold and bold italic indicate top 1st and 2nd rank, respectively.

| Datasets | Rain200L | Rain200H | Rain1400 | SPA-Data |
|---|---|---|---|---|
| Metrics | PSNR/SSIM | | | |
| Input | 26.70/0.8438 | 13.07/0.3733 | 25.24/0.8097 | 34.15/0.9269 |
| RESCAN | 37.07/0.9867 | 26.60/0.8974 | 32.03/0.9314 | 38.11/0.9707 |
| SPANet | 31.59/0.9652 | 23.04/0.8522 | 29.85/0.9148 | 40.24/0.9811 |
| SIRR | 30.04/0.9360 | 21.25/0.7639 | 28.44/0.8893 | 35.31/0.9411 |
| DRD-Net | *37.15/0.9873* | *28.16/0.9201* | 32.57/0.9388 | -/- |
| RCDNet | 35.28/0.9712 | 26.18/0.8356 | *33.04/0.9472* | *41.47/0.9834* |
| **Ours** | **38.54/0.9880** | **28.35/0.9217** | **33.51/0.9549** | **42.72/0.9876** |

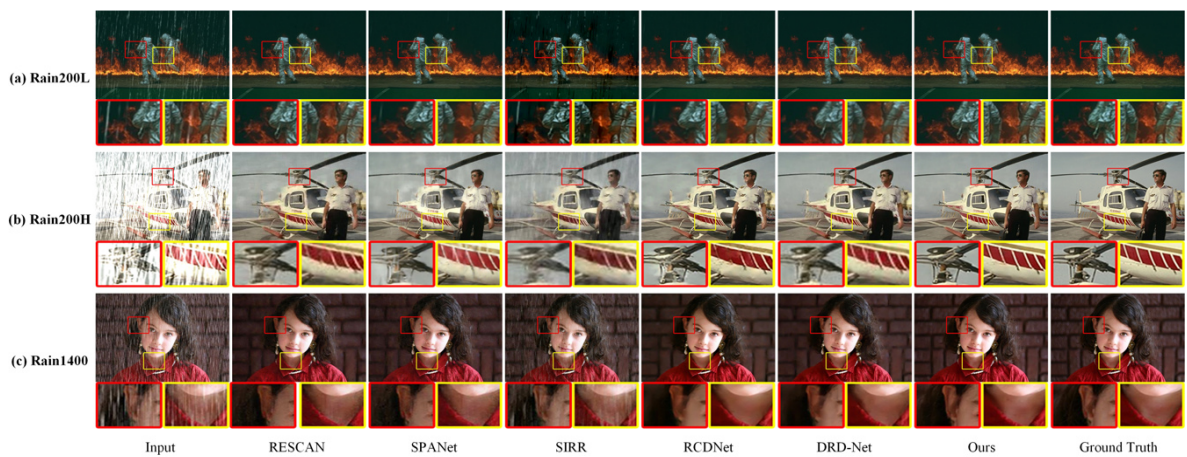

Figure 5. Visual comparison of three synthetic examples, including (a) Rain200L, (b) Rain200H, and (c) Rain1400. Zooming in the figures offer a better view at the deraining capability.

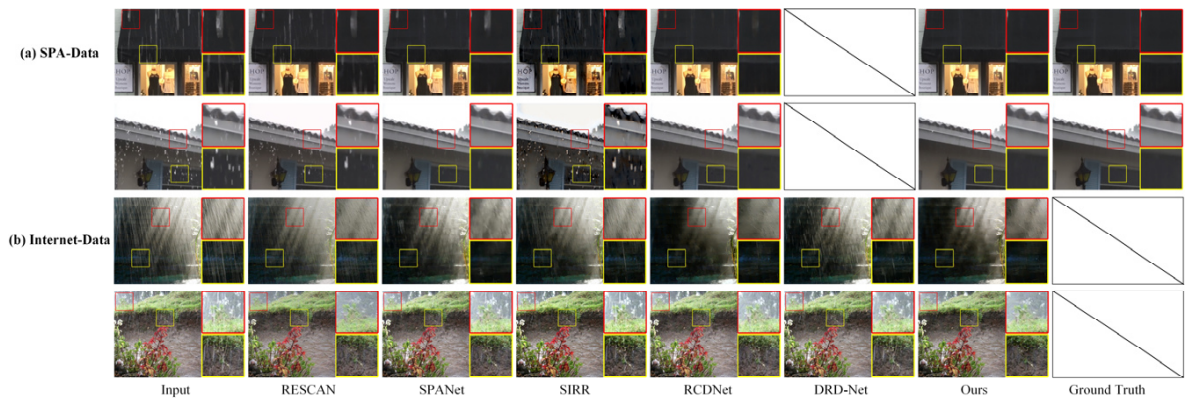

Figure 6. Visual comparison of all the competing methods on two real-world datasets, including SPA-Data (a) and Internet-Data (b). Zooming in the figures offer a better view at the deraining capability.

### 4.3. Ablation Studies

We conduct the ablation analysis to verify the effectiveness of parameters and configurations. All the following studies carry out in the same platform using the Rain200H dataset for training process.

**Multi-scale Hourglass Extractive Block Numbers.** To explore the number impacts on feature extraction, we carry out the deraining experiments with different numbers of MHEB to the designed network. In detail, the numbers of MHEB are set to $N \in \{4,6,8,10\}$ and the corresponding PSNR/SSIM results can be illustrated in Table 3. As display, the increasing blocks can generate higher PSNR value, leading to a better extractive performance. The PSNR improvement seem limited after N=8 with massive calculated cost. Therefore, we select N=8 as the default parameter to balance the extraction effects and computational complexity.

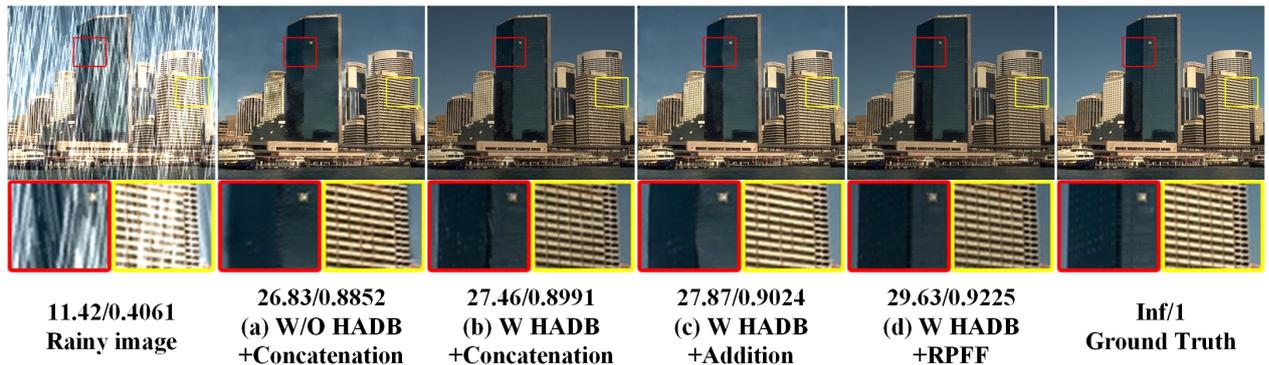

Figure 7. Visual and quantitative comparison of different models, with the description of PSNR/SSIM under the images.

Table 3. Ablation study on Multi-scale Hourglass Extractive Block (MHEB) numbers.

| Block No. | N=4 | N=6 | N=8 (default) | N=10 |
|---|---|---|---|---|
| PSNR/SSIM | 27.28/0.9122 | 27.96/0.9180 | **28.35/0.9217** | 28.37/0.9214 |

**Effect of feature distillation.** Taking 8 MHEBs as an example, the model with HADB needs only 70% percent of network parameters in contrast to the one without HADB. As the network grow deeper, HADB model can further help to eliminate the useless features. For fair comparison, we use the simple concatenated operation to verify the distillation effect. As can be observed in Figure 7(a), the model without HADB makes less contribution in preserving the spatial and structure details. When integrating HADB in Figure 7(b), the derained result achieves a better performance in restoring the detail information. In summary, our HADB can enhance detail restoration with fewer parameters and computation cost.

**Effect of feature fusion strategy.** We further analysis our RPFF effect by comparing with other two commonly used fusion operators, e.g. feature addition and concatenation. Figure 7(b-d) present the vision and evaluation criteria of mentioned feature fusion strategies. As display in Figure 7(b), the derained result of concatenation strategy leaves some rain streaks and detail loss. Figure 7(c) shows the addition strategy, which has less rain streaks existing but contain more content blurring. Compare to feature addition and concatenation, our RPFF do a better performance in rain streaks removal, detail restoration and less content blurring. The derained results demonstrate the effectiveness of the designed fusion module in progressively aggregating features for single image deraining.

## 5. Conclusion

In this paper, we propose a Multi-scale Hourglass Hierarchical Fusion Network (MH$^2$F-Net) to deal with the single image deraining. A novel multi-scale hourglass structure is designed to extract local and global features at different scales. Specially, a hierarchical attentive distillation block is first involved to recalibrate the hierarchical features by using dual attention feature responses. In addition, an advanced residual projected feature fusion strategy is introduced to achieve comprehensive feature aggregation, so that the features from different sources are progressively discriminated and fused for improving deraining performance. Quantitative and visual results demonstrate our developed model outperforms other comparing deraining approaches on both synthetic and real-world rainy datasets.